\documentclass[aps,preprint,nofootinbib,floatfix,superscriptaddress]{revtex4}
\usepackage{color}
\usepackage{slashed}
\usepackage{amsmath,amssymb,graphicx,xcolor,mathtools,float}
\usepackage{graphicx}
\usepackage{epstopdf}
\usepackage[colorlinks=true,linktocpage=true,linkcolor=red,citecolor=blue]{hyperref}
\newcommand{\nn}{\nonumber}
\newcommand{\be}{\begin{equation}}
\newcommand{\ee}{\end{equation}}
\newcommand{\bea}{\begin{eqnarray}}
\newcommand{\eea}{\end{eqnarray}}

\newcommand{\ep}{\epsilon}
\newcommand{\om}{\omega}  
\newcommand{\vp}{\vec p}
\newcommand{\del}{\partial}

\newcommand{\diracslash}[1]{#1\llap{/\kern2pt}}

\def\be{\begin{equation}}
\def\ee{\end{equation}}
\def\bearr{\begin{eqnarray}}
\def\eearr{\end{eqnarray}}

\def\vec#1{\mathchoice
        {\mbox{\boldmath $#1$}}
        {\mbox{\boldmath $#1$}}
        {\mbox{\boldmath $\scriptstyle #1$}}
        {\mbox{\boldmath $\scriptscriptstyle #1$}}
}
\begin{document}

\title{Anisotropic electrical conductivity of magnetized hot quark matter}

\author{Aritra Bandyopadhyay}
\affiliation{Departamento de F\'{i}sica, Universidade Federal de Santa Maria, Santa Maria, 
 RS, 97105-900, Brazil}
 \affiliation{Guangdong Provincial Key Laboratory of Nuclear Science, Institute of Quantum Matter, South China Normal University, Guangzhou 510006, China}

\author{Sabyasachi Ghosh}
\affiliation{Indian Institute of Technology Bhilai, GEC Campus, Sejbahar, Raipur 492015, 
Chhattisgarh, India} 
 
\author{Ricardo L. S. Farias}
\affiliation{Departamento de F\'{i}sica, Universidade Federal de Santa Maria, Santa Maria, 
RS, 97105-900, Brazil}

\author{Jayanta Dey}
\affiliation{Indian Institute of Technology Bhilai, GEC Campus, Sejbahar, Raipur 492015, 
Chhattisgarh, India}

\author{Gast\~ao Krein}
\affiliation{Instituto de F\'{i}sica Te\'orica, Universidade Estadual Paulista,  
Rua Dr. Bento Teobaldo Ferraz, 271 - Bloco II, 01140-070 S\~ao Paulo, SP, Brazil}

\begin{abstract}
\noindent
We studied the effect of a strong magnetic field ($B$) on the electrical conductivity of hot quark matter.
The electrical conductivity is a key transport coefficient determining the time dependence
and strength of magnetic fields generated in a relativistic heavy-ion collision. A~magnetic field 
induces Hall anisotropic conduction, phase-space Landau-level quantization and, if sufficiently strong, 
interferes with prominent QCD phenomena such as dynamical quark mass generation, likely affecting the quark 
matter electrical conductivity, which depends strongly on the quark masses. To address these issues, we 
used a quasi-particle description of quark matter in which the electric charge carriers are 
constituent quarks with temperature- and magnetic-field-dependent masses predicted by a Nambu--Jona-Lasinio 
model. The model accurately describes recent lattice QCD results showing magnetic catalysis at low temperatures 
and inverse magnetic catalysis at temperatures close to the pseudo-critical temperature ($T_{\rm pc}$) of the 
QCD phase transition. We found that the magnetic field increases the conductivity component parallel to it and 
decreases the transverse component, in qualitative agreement with recent lattice QCD results. In addition, we 
found that: (1)~the space anisotropy of the conductivity increases with~$B$, (2)~the longitudinal conductivity  
increases due to phase-space Landau-level quantization, (3)~a lowest Landau level approximation behaves poorly 
for temperatures close to $T_{\rm pc}$, and (5)~inverse magnetic catalysis leaves a distinctive signal in all 
components of the conductivity, a prominent peak at $T_{\rm pc}$. 
Our study adds to the existing body of work on the hot quark matter electrical conductivity by incorporating 
nontrivial temperature and magnetic field effects on dynamical mass generation. Our results are useful both 
for studies employing magnetohydrodynamics simulations of heavy-ion collisions and for getting insight on 
lattice QCD results.
\end{abstract}
%
%\pacs{12.38.Mh, 21.65.Qr, 12.39.Ki}
%
%\keywords{Quark matter, Transport coefficients, Electrical conductivity, Dynamical chiral symmetry breaking}

\maketitle

%%%%%%%%%%%%%%%%%%%%%%%%%%%%%%%%%%%%%%%%%%%%%%%%%%%%%%%%%%%%%%%%%%%%%%%%%%
%
\section{Introduction} 

Relativistic heavy-ion collisions can produce strong magnetic fields~\cite{Rafelski:1975rf,
Kharzeev:2007jp}. Field strengths comparable to or even larger than the strong-interaction scale 
$\Lambda_{\rm QCD} \simeq 0.25$~GeV can be produced in the collisions. Indeed, field strengths as large 
as $e B \sim 15 m^2_\pi \gg \Lambda^2_{\rm QCD}$ have been estimated~\cite{Skokov:2009qp} 
for Pb-Pb collisions at the Large Hadron Collider. Such strong fields interfere with prominent
strong-interaction phenomena such as dynamical mass generation and the chiral anomaly taking place in 
the hot quark matter{\textemdash{or quark-gluon plasma}. Such interferences 
are predicted to lead to interesting effects, some of which were already observed but the interpretation of the 
observations is ambiguous because alternative explanations are possible{\textemdash}Refs.~\cite{Tuchin:2013ie,
Kharzeev:2015znc,Wang:2016mkm,Zhao:2019hta} are recent reviews on this. An~adversity here is that the fields are 
created early in the collision and weaken fast as the system expands~\cite{Skokov:2009qp,Voronyuk:2011jd}. 
On the other hand, the early-produced fields can induce electric currents in the expanding matter which in turn 
produce magnetic fields that can last while the system exists~\cite{Tuchin:2013apa,Skokov_cBt,
Gursoy:2014aka,Tuchin_cBt}. A~key physical property in the electric-current induction is 
the electrical conductivity, a quantity poorly constrained by the fundamental theory, quantum 
chromodynamics (QCD). 

The quark matter electrical conductivity is poorly constrained by QCD due to two main reasons: the matter
produced in a heavy-ion collision is a strongly-interacting many-body system, and a strong magnetic field modifies 
intrinsic properties of the electric charge carriers in the system. A proper treatment of both features requires 
nonperturbative methods. Lattice QCD, a nonperturbative
first-principles numerical method formulated in Euclidean space, can compute the electrical conductivity{\textemdash}and other 
transport properties such as shear and bulk viscosities{\textemdash}but the computation requires reconstructing spectral functions 
from Euclidean correlation functions using ill-posed inversion methods from imaginary time to frequency space. For $B=0$, the 
first lattices results for the conductivity in full QCD  appeared only recently~\cite{Brandt:2012jc,Amato:2013naa}, although 
results from quenched simulations are available for some time~\cite{Gupta:2003zh,{Aarts:2007wj},{Ding:2010ga}}. 
For $B \neq 0$, there are results from quenched simulations of an SU(2) gauge theory~\cite{Buividovich_cond}
and from simulations in full QCD with $N_f=2+1$ flavors reported in a very recent preprint~\cite{Astrakhantsev:2019zkr}. 
Recent analytical nonperturbative calculations of the conductivity, taking into account effects of the magnetic 
field, have been performed using different phenomenological approaches~\cite{{Nam:2012sg},Hattori_cond1,Hattori_cond2,
Sedarkian_cond,Kerbikov_cond,Feng_cond,{Fukushima:2017lvb},Li:2018ufq,Das:2019wjg,Das:2019ppb}.

Prominent magnetic field effects, however, are neglected in previous studies, namely those that affect intrinsic 
properties of the electric charge carriers in the medium, which are predominantly the light $u$ and $d$ 
quarks. This is an important omission: recent lattice QCD
results~\cite{Bali:2011qj,Bali_PRD,Bruckmann:2013oba,Endrodi:2013cs} have shown that strong magnetic fields have 
dramatic effects on the QCD phase diagram, notably in the region close to the pseudo-critical temperature $T_{\rm pc} 
\simeq 0.170$~GeV, the region associated with the hadron-to-quark transition. The effects are most striking on the $u$ 
and $d$ quark condensates, namely the condensates increase with the magnetic field for low temperatures and decrease for 
temperatures close to $T_{\rm pc}$; in the first case one refers to magnetic catalysis (MC) and in the latter to inverse 
magnetic catalysis (IMC){\textemdash}Refs.~\cite{Marco2,Miransky,jens_rmp} are recent reviews which contain extensive 
lists of references on this subject. The quark condensates have a direct impact on the effective masses of 
the $u$ and $d$ quarks, which in turn play an important role in the conductivity. 

In the present work we fill this gap in the study of the electrical conductivity of magnetized quark matter
by using a quasi-particle model. Much of our understanding of the low-energy regime of QCD, and of the QCD phase diagram 
in particular, is built on insights gained with quasi-particle models. Among the several existing quasi-particle models, 
those based on the Nambu--Jona-Lasinio (NJL) model~\cite{Nambu:1961tp,Nambu:1961fr} have been valuable in this 
respect{\textemdash}Refs.~\cite{Vogl:1991qt,Klevansky:1992qe,Hatsuda:1994pi,Buballa:2003qv} are extensive reviews 
on the model in different QCD applications. In these models, the primary electric charge carriers 
are the $u$ and $d$ quarks and their response to electromagnetic forces are strongly dependent on their 
effective in-medium masses. The effective quark masses are determined by the in-medium quark condensate. As mentioned,
strong magnetic fields change the condensate and in the present paper take this effect into account in the computation
of the electrical conductivity. We use the NJL model of Refs.~\cite{Farias:2014eca,Farias:2016gmy},
a model that reproduces the lattice QCD data for the quark condensates, showing MC at low 
temperatures and IMC at temperatures close to $T_{\rm pc}$. We derive the expressions of the components 
of the conductivity by solving the Boltzmann equation in the relaxation-time approximation.

For a realistic quantification of magnetic fields effects in a heavy-ion collision, it is imperative 
to take into account the temperature and magnetic field dependence of the electrical conductivity. Simulations of
the field dynamics invariably involve solving relativistic magnetohydrodynamics equations. These equations also require 
other transport coefficients, as shear and bulk viscosities. The magnetic field dependence 
of the shear viscosity was computed recently in Refs.~\cite{Li:2017tgi,Nam_shear,Sedarkian_shear,Tawfik_shear,
Tuchin_shear,G_shear_NJLB,Mohanty:2018eja,JD1,JD2,HRGB} and of the bulk viscosity in Refs.~\cite{Hattori:2017qih,Sedarkian_bulk,
Huang_bulk,Agasian_bulk1,Agasian_bulk2}. Our study of the electrical conductivity adds to this body of work by
incorporating important effects neglected in previous studies. Our results should be useful 
for studies employing magnetohydrodynamics simulations of heavy-ion collisions and for getting insight
on lattice QCD results.

The paper is organized as follows. In Sec.~\ref{sec:formalism} we start reviewing the NJL model quasi-particle 
description of magnetized quark matter, with particular emphasis on the magnetic catalysis and inverse magnetic
catalysis of the chiral condensate. Then we present the derivation of the electrical conductivity by solving
the Boltzmann equation in the relaxation time approximation. In Sec.~\ref{sec:results} we present numerical 
results for the conductivity and discuss their meaning. In particular, we discuss the implications of the changes 
in the quark masses on two well known magnetic field effects on the conductivity: breaking of the space isotropy 
of the conductivity, and dimensional reduction of the dynamics along with quantization of phase space (Landau levels). 
The first happens when mutually perpendicular electric and magnetic fields exist in the conducting medium, field 
configurations that actually can be generated in a heavy-ion collision~\cite{Tuchin:2013ie,Tuchin:2013apa}. The 
second is a quantum mechanical effect that is particularly important for strong fields. Section~\ref{sec:summary} 
presents a summary and the perspectives of our study.

%%%%%%%%%%%%%%%%%%%%%%%%%%%%%%%%%%%%%%%%%%%%%%%%%%%%%%%%%%%%%%%%%%%%%%%%%%
%
\section{Formalism}
\label{sec:formalism}

We start with a brief review of the NJL model quasi-particle description of quark matter 
at finite temperature~$T$ and in the presence of a magnetic field with strength~$B$. We 
focus on the implications of the inverse magnetic catalysis (IMC) phenomenon on the quasi-particle effective mass~$M$, 
the constituent quark mass. Next, we present the derivation of the electrical conductivity within 
the relaxation time approximation. The main input from the 
NJL model in these derivations is the $T-$ and $B-$dependent constituent quark mass~$M$. 

% % % % % % % % % % 
%
\subsection{Quark matter in presence of a magnetic field}
\label{Sec:NJLB}

The Lagrangian density for the isospin-symmetric two-flavor version of NJL model 
in presence of an electromagnetic field ($A^\mu$) is given by
\be
\mathcal{L}_{NJL}= -\frac{1}{4} F^{\mu\nu}F_{\mu\nu}
+ \bar{\psi}\left(\slashed{D}-m\right)\psi
+ G\left[ (\bar{\psi}\psi)^2+(\bar{\psi}i\gamma_5{\vec\tau}\psi)^2\right],
\label{NJL_lag}
\ee
where $\psi$ is a flavor doublet of $u$ and $d$ quark fields, each being an $N_c-$plet, where
$N_c = 3$ is the number of colors, $m = {\rm diag}(u,d)$ the quark-mass matrix, 
$D_\mu = i\partial_\mu - Q A_\mu$ the covariant derivative, 
$Q = {\rm diag}(q_u = 2e/3, q_d =-e/3)$ the charge matrix, $A_\mu$ the electromagnetic gauge field,
$F_{\mu\nu} = \partial_\mu A_\nu - \partial_\nu A_\mu$, and $\vec \tau = (\tau^1, \tau^2, \tau^3)$ 
are the isospin Pauli matrices. We work in the approximation of exact isospin symmetry, i.e. $m_u = m_d$. 
The model is solved in the quasi-particle approximation or, equivalently, in the mean-field
approximation, which corresponds to the leading-order approximation in the $1/N_c$ expansion. 
Since the model is unrenormalizable due to the quadratic fermionic interaction, a regularization
procedure must be employed. In the present paper we employ a sharp cutoff $\Lambda$ to
regularize ultraviolet divergences. 

In the quasi-particle approximation, the gap equation for the constituent quark mass $M$ 
at finite temperature $T$ and in the presence of a magnetic field~$B$ is given by
\bea
M = m - 2 G  \sum_{f=u,d}\langle \bar{\psi}_f\psi_f\rangle,
\label{Gap_B}
\eea
where $\langle \bar{\psi}_f\psi_f\rangle$ represents the quark condensate of flavor~$f$, which can be
written as a sum of three contributions~\cite{klimenko,sidney,Farias:2014eca}: 
\bea
\langle \bar{\psi}_f\psi_f\rangle = \langle \bar{\psi}_f\psi_f\rangle^{vac} 
+ \langle \bar{\psi}_f\psi_f\rangle^B + \langle \bar{\psi}_f\psi_f\rangle^{T,B},
\label{gap}
\eea
with 
\bea
\langle \bar{\psi}_f\psi_f\rangle^{vac} &=& -\frac{MN_c}{2\pi^2}\left[\Lambda\sqrt{\Lambda^2+M^2} 
- M^2 \ln \left(\frac{\Lambda+\sqrt{\Lambda^2+M^2}}{M}\right) \right], 
\label{cond1} \\
\langle \bar{\psi}_f\psi_f\rangle^{B} &=& -\frac{M|q_f|BN_c}{2\pi^2}\left[\ln\Gamma(x_f) 
- \frac{1}{2}\ln(2\pi)+x_f-\frac{1}{2}\left(2x_f-1\right)\ln(x_f) \right], 
\label{cond2 } \\
\langle \bar{\psi}_f\psi_f\rangle^{T,B} &=& \sum\limits_{k=0}^\infty \alpha_k \, 
\frac{M|q_f|BN_c}{2\pi^2}\int\limits_{-\infty}^\infty dp_z \, \frac{n_F(\om_f)}{\om_f},
\label{cond3} 
\eea
where $q_f$ is the electric charge of the quark with flavor $f=(u,d)$, 
$\Gamma(x_f)$ is the Euler gamma function, $x_f = {M^2}/{2|q_f|B}$. In addition, in
Eq.~(\ref{cond3}) $k$ indexes Landau levels, with $\alpha_k = 2 - \delta_{k,0}$  being 
spin degeneracy, and $n_F(\om_f)$ is the Fermi-Dirac distribution function:
\be
n_F (\om_f) = \frac{1}{1 + e^{\beta \om_f}},
\label{FD-dist}
\ee
where 
\be
\om_f = (p_z^2+M^2+2k|q_f|B)^{1/2}.
\label{omega_f}
\ee
Notice that, contrary to the vacuum case, in the presence of a magnetic field the 
$u$~and~$d$ condensates are different due to the different $u$~and~$d$ electric charges, 
but the constituent masses of the $u$ and $d$ quarks are still equal, i.e. $M$ stands
for both $M_u$ and $M_d$. 

As mentioned earlier, the NJL model in the quasi-particle approximation is unable to describe inverse 
magnetic catalysis (IMC)~\cite{Bali:2011qj,Bali_PRD}, unless one imposes that the coupling constant~$G$ 
of the model is $T-$ and $B-$dependent~\cite{Farias:2014eca}. A precise description of the lattice results 
for the $u$ and $d$ quark condensates is obtained within the NJL model with the 
parametrization~\cite{Farias:2016gmy}:
\bea
G(eB,T) = c(eB)\left[1-\frac{1}{1+e^{\beta(eB)[T_a(eB)-T]}} \right]+s(eB),
\label{GBT}
\eea
where $c(eB)$, $\beta(eB)$, $T_a(eB)$ and $s(eB)$ depend only on the magnitude of the external magnetic 
field. Their values for selected values of $B$ are given in Table~1 of Ref.~\cite{Farias:2016gmy}. All numerical
results presented here refer to this parametrization. 

The model just described was used to study the effects of a magnetic field on neutral pions in Refs.~\cite{{Avancini:2016fgq},
Avancini:2018svs}, with results that agree with corresponding lattice QCD results~\cite{{Bali:2015vua}}. 
Further evidence for the association of the IMC phenomenon with a magnetic field 
dependence decreasing the coupling strengths of quark-matter effective degrees of freedom was 
given in Refs.~\cite{Ayala_LSM1,Ayala_LSM2}.

% % % % % % % % % % 
%
\subsection{Electrical conductivity in presence of magnetic field: classical and quantum}
\label{sec:el_B}

Here we derive the electrical conductivity of magnetized hot quark matter (composed by the constituent
quarks and antiquarks of flavor $u$ and $d$) employing the relaxation time approximation{\textemdash}
closely following the approach of Ref.~\cite{Sedarkian_cond}. Initially we consider the classical description, ignoring 
the quantization of phase space in terms of Landau levels.

Let us consider a relativistic electrically charged fluid of 
constituent quarks (and antiquarks) with masses $M_f$ and energies $\omega_f$ (defined 
in Eq.~(\ref{omega_f})), characterized by distribution functions $f(\om_f)$ described 
by the Boltzmann equation. In the electric-charge-transport picture, an external electric 
field is responsible for driving the system out of equilibrium. Let~$f_0(\om_f) = n_F(\om_f)$ 
denote the equilibrium distribution, where $n_F(\om_f)$ is the Fermi-Dirac distribution given in 
Eq.~(\ref{FD-dist}), and $f(\om_f) = f_0(\om_f) + \delta f(\om_f)$ the distribution in the presence the 
electric force. The electric current $\vec J_f$ associated with a given quark flavor $f$ is expressed 
in terms of the quark 
distribution function~by
\bea
\vec J_f &=& \vec J_{0 f} +  \vec J_{\delta f} \nn \\
&=& q_f \, g \int \frac{d^3p}{(2\pi)^3} \, \vec v_f \left[ f_0(\omega_f) + \delta f(\omega_f)\right],
\label{Current_Kinetic}
\eea
where $q_f$ is the charge of the quark (or antiquark), $g$ a spin-color degeneracy factor, 
and $\vec v_f = {\vp}/{\om_f}$ the quark velocity. The electrical conductivity of the fluid 
is a tensor relating the induced electric current and the electric field. Since we are working 
in the isospin-symmetric limit, we can drop the flavor index~$f$ and concentrate on the 
contribution of a given quark-flavor~$f$ to the conductivity and add up the contributions of 
each flavor at the end; the only difference comes from the electric charge $q_f$. From
Ohm's law we have 
\be
J^i_{\delta} = \sigma^{ij} \, E^j .
\label{Ohm}
\ee

The Boltzmann equation for the distribution function $f(\om)$ of a quark of a given flavor 
under the influence of a generic external force $\vec F$ is given by
\be
\frac{\partial f}{\partial t} + \vec v \cdot \frac{\partial f}{\partial \vec x} 
+ \vec F \cdot \frac{\partial f}{\partial \vp} 
= \left(\frac{\partial f}{\partial t}\right)_{\rm coll},
\label{Boltz_B0}
\ee
where $(\partial f/\partial t)_{\rm coll}$ 
takes into account the collisional effects. For an uniformly 
distributed system, ${\partial f}/{\partial \vec x}=0$. The force $\vec F$
drives the system out of equilibrium and when the force is removed, scattering 
events described by $(\partial f/\partial t)_{\rm coll}$ will restore equilibrium. 
The relaxation time approximation (RTA) consists in assuming for the collisional term
\be
\left(\frac{\partial f}{\partial t}\right)_{\rm coll} = - \frac{\delta f}{\tau_c},
\label{RTA}
\ee
where $\tau_c$ is the relaxation time, the time required for 
the system to return to equilibrium after removing $\vec F${\footnote{Indeed, for $\vec F = 0$
the solution of Eq.~(\ref{Boltz_B0}) in the RTA is given by
\be
f(t) = f_0 + [f(t_0) - f_0] e^{-(t-t_0)/\tau_c},
\ee
where $t_0$ is the time at which $\vec{F}$ is removed.}}. The determination of the 
$T$ and $B$ dependence of $\tau_c$ is out of the scope of the present paper; when presenting 
results in the next section, we use values for $\tau_c$ in the range $0.2 - 10$~fm/c, the 
latter number being approximately the lifetime of the medium produced at the RHIC and the LHC. 
To assess the impact of a $T$ and $B$ dependence of $\tau_c$ on the results, we use a $\tau_c$ 
computed along the lines of the approach introduced in Ref.~\cite{G_shear_NJLB}, based on 
quark-quark scattering cross-sections calculated with
the contact interactions of the NJL model. The validity of the calculation is, however, 
restricted to $B \ll 10 m^2_\pi$ and therefore we present results for low values of $B$ only.

In the present context, $\vec F$ is the Lorentz's force, $\vec F = q (\vec E + \vec v \times \vec B)$,
where $q$ is the electric charge (recall that we are suppressing the flavor index~$f$). It is 
useful to discuss first the situation with $\vec B = 0$; the force is then $\vec F = q \vec E$. 
Assuming $\delta f \ll f_0$, Eq.~(\ref{Boltz_B0}) implies
\bea
\delta f &=& - \tau_c \, q {\vec E} \cdot \frac{\partial f_0(\om)}{\partial \vp} \nn \\
&=& \tau_c \, \beta \, q \, \frac{\vp \cdot \vec E}{\omega} \, f_0(\om) [1 - f_0(\om)] ,
\label{del_f}
\eea
where $\beta = 1/T$.  Using this into Eq.~(\ref{Ohm}), adding the contributions from quarks and antiquarks 
of flavor $u$ and $d$, taking into account that the spin-color degeneracy factor is $g=2 \times 3$ and
the sum over flavor gives $\sum_f q^2_f = {5e^2}/{9}$, : for the conductivity
\be
\sigma^{ij} = \delta^{ij} \, \sigma,
\ee
with 
\be
\sigma = e^2 \beta \, \frac{20}{9}  \int \frac{d^3p}{(2\pi)^3} \, 
\frac{\vp^2}{\om^2} \, \tau_c \, f_0(\om) [1 - f_0(\om)] .
\label{sigma_B0}
\ee

Next, we proceed with the derivation of $\sigma^{ij}$ in presence of a magnetic field 
$\vec B$ of arbitrary strength. In this situation, it is necessary to include the $\delta f$
contribution in the drift term in Eq.~(\ref{Boltz_B0}); otherwise, the term $\vec v \times \vec B$
in the Lorentz gives null contribution, because:
\be
\left(\vec v \times {\vec B}\right)\cdot \frac{\partial f_0(\om)}{\partial \vp} =
\left(\frac{\vp}{\om} \times \vec B\right)\cdot\vp \,\frac{\del f_0(\om)}{\del\om} = 0.
\ee
Including the $\delta f$ contribution in the drift term in Eq.~(\ref{Boltz_B0}), one
obtains
\bea
\delta f &=& - \tau_c q \left[ \frac{1}{\om} \, \vp \cdot {\vec E}\, 
\frac{\partial f_0}{\partial \om} 
+ \left(\frac{\vp}{\om}\times {\vec B}\right)\cdot \frac{\partial (\delta f)}{\partial \vp}\right] .
\label{RBE_H_df}
\eea
Since any term proportional to $\vp$ from $\partial(\delta f)/{\partial \vp}$ 
leads to a vanishing contribution, one must have $\delta f \sim \vp \cdot \vec {\cal F}$ 
times a function of $\vp^2$, where $\vec{\cal F}$ is a vector that depends on the $\vec E$ and $\vec B$ 
fields. As shown in Ref.~\cite{Sedarkian_cond}, Eq.~(\ref{RBE_H_df}) can be solved by writing 
$\delta f = - \vp \cdot \vec {\cal F}\; \partial f_0/\partial \om$, with
$\vec{\cal F} = \alpha \vec{e} + \beta \vec{b} 
+ \gamma \vec{e}\times \vec{b}$, where $\vec{e} = {\vec E}/E$ and 
$\vec{b} = {\vec B}/B$, where $E = |\vec E|$ and $B=|\vec B|$. Specifically:
\be
- \vp \cdot \vec {\cal F} = - \tau_c \, q \left[ \frac{1}{\om} \, \vp \cdot {\vec E}
- \frac{1}{\om} \vp \cdot \left(\vec B \times \vec{\cal F}\right) \right],
\ee 
from which one obtains 
\bea
\alpha &=& q \left(\frac{\tau_c}{\om}\right)\frac{1}{1+(\tau_c/\tau_B)^2}\, E,
\label{alpha}
\\
\beta &=& q \left(\frac{\tau_c}{\om}\right)\frac{(\tau_c/\tau_B)^2}{1+(\tau_c/\tau_B)^2}
\, (\vec {e}\cdot \vec{b}) E,
\label{beta}
\\
\gamma &=& \left(\tau_c/\tau_B \right) \, \alpha 
= q \left(\frac{\tau_c}{\om}\right) \frac{(\tau_c/\tau_B)}{1+(\tau_c/\tau_B)^2}
\, E,
\label{gamma}
\eea
where $\tau_B = \om/qB$ is the inverse of the cyclotron frequency. Therefore, one 
obtains for $\delta f$:
\be
\delta f = - q \left(\frac{\tau_c}{\om}\right) \frac{1}{1+(\tau_c/\tau_B)^2}
\left[\delta_{ij}
+ (\tau_c/\tau_B)\ep^{ijk}b^k+(\tau_c/\tau_B)^2 b^i b^j\right] \frac{\partial f_0(\om)}{\partial \om} 
\, p^i E^j .
\label{phi_Ej}
\ee
From Ohm's law in Eq.~(\ref{Ohm}) and taking flavor summation of each quark or anti-quark, having spin and color degeneracy factor $g=2 \times 3$, one immediately obtains 
for $\sigma^{ij}$:
\be
\sigma^{ij} = \delta^{ij} \, \sigma_0 + \ep^{ijk}b^k \, \sigma_1 + b^i b^j\,\sigma_2,
\ee
where $\sigma_n$, $n=0,1,2$, are given by
\bea
\sigma_n &=&  2\beta\sum_{f=u,d} q_f^2   \int \frac{d^3p}{(2\pi)^3} \, 
\frac{\vp^2}{\om^2} \,  
\frac{\tau_c \, (\tau_c/\tau_{B,f})^n}{1+(\tau_c/\tau_{B,f})^2}  \, f_0(\om) [1 - f_0(\om)]  
\label{cond_ne1} \\[0.3true cm]
&=&   2\beta\sum_{f=u,d} q_f^2 \int \frac{d^3p}{(2\pi)^3} \, 
\frac{\vp^2}{\om^2} \, \frac{ \tau_c \, (\tau_{B,f}/\tau_c)^{2-n}}{1+(\tau_{B,f}/\tau_c)^2}   
\, f_0(\om)[1 - f_0(\om)]~,
\label{cond_ne2} 
\eea 
where flavor index is marked in charge ($q_f$) as well as in cyclotron time period ($\tau_{B,f}$). 
These expressions are valid for quarks and antiquarks. Since $\sigma_0$ and $\sigma_2$ 
are proportional to the square of the electric charge $q_f$ and the medium we are considering contains 
equal numbers of quarks and antiquarks (zero chemical potential, $\mu =0$), the total 
$\sigma_0$ and $\sigma_2$ of the plasma is obtained by adding the quark and antiquark contributions. It means that additionally factor 2 will be multiplied in $\sigma_0$ and $\sigma_2$ of Eq.~(\ref{cond_ne2}). This will not hold for $\sigma_1$. The quark and antiquark contributions to $\sigma_1$ have opposite sign due to the
odd-function dependence of $\tau_B=\om/(q_f B)$ or $q_f$; therefore, the total Hall conductivity of the plasma is zero.

When $\vec B$ is in the $z$ direction, which implies $\hat b^z = 1$ and $\hat b^x = \hat b^y = 0$,  
the conductivity matrix elements are given by: $\sigma^{zz} = \sigma_0 +\sigma_2 = \sigma$ (longitudinal
conductivity), $\sigma^{xx} = \sigma^{yy} = \sigma_0$ (transverse conductivity), and
$\sigma^{xy} = -\sigma^{yx} = \sigma_1$ (Hall conductivity), 
with the other components being zero. The  $\sigma^{xy} = - \sigma^{xy} = \sigma_1$ components are absent
in the zero magnetic field case, Eq.~(\ref{sigma_B0}). This reveals that the space-anisotropic
nature of the conductivity which, according to Eqs.~(\ref{cond_ne1}) and (\ref{cond_ne2}), is 
controlled by the ratio between the two time scales $\tau_c$ and $\tau_B$. This anisotropy shows 
interesting features for weak and strong magnetic fields.

The two-way expression of the $\sigma_n$ components in Eqs.~(\ref{cond_ne1}) and 
(\ref{cond_ne2}) with respect to the $\tau_c$ and $\tau_B$ dependence is useful for discussing 
the weak- and strong-field limits of the $\sigma_n$. For a weak magnetic field,  $\tau_c/\tau_B$ is 
small. Expanding the integrand Eq.~(\ref{cond_ne1}) up to 
second order in $\tau_c/\tau_B$ one obtains:
\be
\left(\begin{array}{c}
\sigma_0 \\
\sigma_1 \\
\sigma_2 \end{array}
\right)
\simeq  
2\beta \sum_{f=u,d} q_f^2   \int \frac{d^3p}{(2\pi)^3} \, 
\frac{\vp^2}{\om^2} \, \tau_c 
\left(\begin{array}{c}
1-(\tau_c/\tau_{B,f})^2 \\
\tau_c/\tau_{B,f} \\
(\tau_c/\tau_{B,f})^2 \end{array}
\right) \, f_0(\om) [1 - f_0(\om)].
\label{classical}
\ee
This shows that a small space-anisotropic conduction is already present in the  weak-field limit. The 
$\sigma^{xx}$ and $\sigma^{yy}$ diagonal elements are reduced from the zero-field case, whereas the $\sigma^{zz} 
= \sigma_0 + \sigma_2 = \sigma$ component is unchanged. For a strong magnetic 
field, $\tau_c/\tau_B$ is large and Eq.~(\ref{cond_ne2}) up to second order in $\tau_B/\tau_c$ can be written
as
\be
\sigma_n \simeq 2\beta \sum_{f=u,d}q_f^2   \int \frac{d^3p}{(2\pi)^3} \, 
\frac{\vp^2}{\om^2} \,\tau_c  (\tau_{B,f}/\tau_c)^{2-n} \, f_0(\om) [1 - f_0(\om)].
\label{sig_cl}
\ee
One sees that as $B \rightarrow \infty$, all the components of the conductivity tensor, with the 
exception of $\sigma^{zz}$, vanish.

Up to now, classical physics was used to determine the conductivity. As seen in Sec.~(\ref{Sec:NJLB}), 
a proper quantum mechanical treatment of electric charge dynamics in a magnetic field leads to one-dimensional dynamics 
and energy quantization in terms of Landau levels. In practical terms, these effects amount to modifying the 
results above by the following replacements:
\bea
\om = (\vp^2+M^2)^{1/2} &~~~\rightarrow~~~&  \om_{f,k} = (p_z^2+M^2+2k|q_f|B)^{1/2} ,
\\
 3\times 2 \int \frac{d^3p}{(2\pi)^3}\frac{\vp^2}{3\om^2}&~~~\rightarrow~~~&  3\sum_{k=0}^\infty \alpha_k \frac{|q|B}{2\pi} 
\int\limits^{+\infty}_{-\infty} \frac{dp_z}{2\pi}\frac{p_z^2}{\om^2} ,
\label{CM_QM}
\eea
where in the last line we used the fact that the lowest Landau level (LLL), $k=0$, is 
spin-polarized{\textemdash}recall that $\alpha_k = 2 - \delta_{k,0}$. In quantum picture energy will depend on Landau level $k$ and flavor $f$ both, therefore, it is marked as $\om_{f,k}$.
Hence, using those changes, the 
longitudinal conductivity of $u/d$ flavor quark/antiquark in the quantum description is given by 
\bea
\sigma^{zz} = 3 \beta \sum_{f=u,d} q_f^2   \sum_{k=0}^\infty \alpha_k \frac{|q_f|B}{2\pi} 
\int\limits^{+\infty}_{-\infty} \frac{dp_z}{2\pi} \frac{{p_z}^2}{\om^2_{f,k}} \tau_c f_0(\om_{f,k})[1-f_0(\om_{f,k})]~.
\label{Lsig_QM_u}
\eea
Considering charges of $u$ and $d$ quarks and their antiparticle contribution, Eq.~(\ref{Lsig_QM_u}) will
be modified to
\bea
\sigma^{zz} &=&6 \beta \sum_{f=u,d} q_f^2   \sum_{k=0}^\infty \alpha_k \frac{|q_f|B}{2\pi} 
\int\limits^{+\infty}_{-\infty} \frac{dp_z}{2\pi} \frac{{p_z}^2}{\om^2_{f,k}} \tau_c f_0(\om_{f,k})[1-f_0(\om_{f,k})]~.
\label{Lsig_QM}
\eea
%
%where the factor~2 comes from summing over the quark and antiquark contributions. 
The transition between 
the classical and quantum regimes in $\sigma^{zz}$ can be obtained by varying $\tau_c$ for a fixed $\tau_B$. 
This analysis is made in the next section.

In the strong field limit, the dominant contribution to $\sigma^{zz}$ comes from the lowest Landau level (LLL). At LLL, energy expression will be transformed as
\be
\om_{f,k=0}=\om_0=\sqrt{p_z^2+M^2}
\ee
and flavor summation can be done as
% because the energy $\om_{f,k}$, which appears in the denominator in Eq.~(\ref{Lsig_QM}) and in the Fermi-Dirac distributions, 
% increases with~$B$ for $k\neq0$. 
%
\bea
\sigma^{zz} &=& 2 \times 3 \left[\left(\frac{2}{3}\right)^3 +
\left(\frac{1}{3}\right)^3\right] e^2 \beta  \sum_{k=0}^\infty \alpha_k \frac{|e|B}{2\pi} 
\int\limits^{+\infty}_{-\infty} \frac{dp_z}{2\pi} \frac{{p_z}^2}{\om^2_{0}} \tau_c f_0(\om_0)[1-f_0(\om_0)]
\nn\\
&=& e^2\beta \frac{|e|B}{\pi} 
2\int\limits^{\infty}_{0} \frac{dp_z}{2\pi} \frac{{p_z}^2}{\om^2_{0}} \tau_c f_0(\om_0)[1-f_0(\om_0)]~
\nn\\
&=& e^2\beta \frac{|e|B}{\pi^2} 
\int\limits_M^\infty d\om_0 \frac{\sqrt{\om_0^2-M^2}}{\om_0} \tau_c f_0(\om_0)[1-f_0(\om_0)]~.
\label{Lsig_LLL}
\eea
%
% \bea
% \sigma^{zz} &=& 3 \sum_f q_f^2 \beta  \frac{|q_f|B}{2\pi^2} 
% \int\limits^{+\infty}_{-\infty} dp_z \frac{{p_z}^2}{\om_0^2} \tau_c f_0(\om)[1-f_0(\om)]~
% \nn\\
% &=& 3 \times 2 \sum_f q_f^2 \beta   \frac{|q_f|B}{2\pi^2} 
% \int\limits_M^\infty d\om_0 \frac{\sqrt{\om_0^2-M^2}}{\om_0} \tau_c f_0(\om)[1-f_0(\om)]~,
% \label{Lsig_LLL}
% \eea
%

One can check that Eq.~(\ref{Lsig_LLL}) coincides with Eq.~(D.5) of Ref.~\cite{Hattori_cond1}.
Eq.~(\ref{Lsig_QM}) contains quantum effects leading to Landau-level quantization, 
but does not contain contributions due to the quantum chiral anomaly. The latter causes 
an imbalance between left- and right-handed quark and antiquark currents and can affect the 
longitudinal electrical conductivity, with the effect increasing with~$B$~\cite{Son:2012bg}.
Such an effect on the $T$ and $B$ dependence on the conductivity has not yet been considered 
in the present context and we reserve for a future publication examination of such effects.

%{\color{red}It seems we had an error in the previous derivations: the source of the error was on coming
%from the first to the second line in Eq.~(\ref{Lsig_LLL}). The integrand in the first line is even in $p_z$, then
%one can write  $\int\limits^{+\infty}_{-\infty} dp_z  \, \cdots = 2 \int\limits^{+\infty}_{0} dp_z \, \cdots$ and in the 
%sequence make the change of variable $p_z \rightarrow \om_0$.} {\color{blue}This requires redrawing the LLL curves
%in Fig.~\ref{elzz}, but nothing terribly important will happen because a factor 2 in the log scale is barely visible.
%Actually, the factor 2 will improve the LLL approximation, I think. Please check.}

%%%%%%%%%%%%%%%%%%%%%%%%%%%%%%%%%%%%%%%%%%%%%%%%%%%%%%%%%%%%%%%%%%%%%%%%%%
%
\section{Numerical results and discussion}
\label{sec:results}

First, we present results for the constituent quark mass~$M$ as a function of the temperature~$T$ and
the magnetic field~$B$. The quark mass is a key ingredient in the calculation of the electrical 
conductivity; it enters in the definition of quasi-particle energy $\om_{f}$, which in turn enters in 
the definition of cyclotron frequency~$\tau_B$, and in the Fermi-Dirac 
distribution $n_F (\om_f)${\textemdash}it is the link between dynamical chiral symmetry 
breaking and transport properties. Next, we present results for the conductivity using 
temperature and magnetic field independent values of the relaxation time. We start presenting
results for the conductivity components and discuss how its anisotropy is affected by $T$ and 
$B$. We examine the impact of phase-space quantization on the longitudinal component of the 
conductivity and examine the validity of truncating phase space
to the lowest Landau level (LLL). We also discuss how the phenomenon of the inverse magnetic catalysis 
affects the electrical conductivity. We finalize our study examining the impact of using a $T$ and $B$
dependent $\tau_c$ on the longitudinal conductivity. 

The constituent quark mass $M = M(T,B)$ was obtained by solving the gap equation, i.e.  Eq.~(\ref{Gap_B}) for
different values of~$T$ and $B$. As mentioned earlier, when $m_u = m_d = m$, the equality $M_u = M_d = M$ holds both
in vacuum and when $B\neq 0$, but the equality $\langle \bar u u \rangle = \langle \bar d d \rangle$  
holds only in vacuum. The parameters of the model are the current quark mass $m$, the coupling $G$ 
and the cutoff $\Lambda$. Since the interest is to take into account the IMC phenomenon in the phenomenology 
of the electrical conductivity, those parameters were fitted to reproduce the lattice data of Ref.~\cite{Bali_PRD}. 
A precise description of the lattice data for the quark condensate is obtained in~\cite{Farias:2016gmy} when 
using the parametrization given in Eq.~(\ref{GBT}). The values of the parameters $c(B)$, $\beta(B)$, $T_a(B)$, 
and $s(B)$ defining the fitting formula in Eq.~(\ref{GBT}) are displayed in Table~1 of Ref.~\cite{Farias:2016gmy}.
The value of the coupling that reproduces the $T=B=0$  quark condensates of lattice data of Ref.~\cite{Bali_PRD} is
$G = 4.50~{\rm GeV}^{-2}$. The values of the remaining parameters are: $m = 5.5$~MeV and $\Lambda = 650$~MeV and they 
correspond to the vacuum pion decay constant, $f_\pi = 93$~MeV, and vacuum pion mass, 
$m_\pi = 140$~MeV. In the numerical results for the fixed coupling constant we have used $G(0,0) = 4.6311~{\rm GeV}^{-2}$, 
i.e. the value obtained in Eq.(\ref{GBT}) that was fitted to the lattice at high temperature region (more details 
can be found in~\cite{Farias:2016gmy}).
 
Figure~\ref{tB_T}(a) displays the constituent quark mass $M$ as a function of the temperature for zero 
and nonzero magnetic field. For small values of $T$, the mass increases with $B$, 
whereas for temperatures near the pseudo-critical temperature $T_{\rm pc} \simeq 0.17$~GeV it decreases with $B$. 
These features, which are carried over to the $u$ and $d$ condensates, characterize the phenomena of magnetic 
catalysis (MC) at low~$T$ and inverse magnetic catalysis (IMC) at $T \sim T_{\rm pc}$. The impact of these features
on the electrical conductivity, in particular on its anisotropy, will be discussed shortly. 

%%%%%%%%%%%%%%%%%%%%%%%%%%%%%%%%%%%%%%%%%%%%%
\begin{figure}[htb]
\includegraphics[scale=0.3]{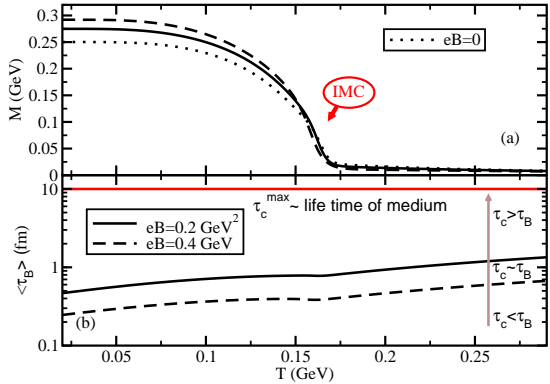} %{tB_T.eps}
\caption{Temperature dependence of (a) constituent quark mass $M$ and (b) momentum average of the cyclotron time 
period $\langle \tau_B \rangle$ (defined in Eq.~(\ref{av-tau})) for different values of the magnetic field. } 
\label{tB_T}
\end{figure}
%%%%%%%%%%%%%%%%%%%%%%%%%%%%%%%%%%%%%%%%%%%%%

The anisotropy is controlled by the ratio of two time scales, the relaxation time $\tau_c$ and the cyclotron 
time $\tau_B = \om/eB$. As mentioned earlier, it is out of our scope to determine $\tau_c$ from microscopic dynamics. We 
recall that $\tau_c$ depends on parameters of the heavy-ion collision as 
the centrality of the collision (impact parameter) and the energy and size of the colliding nuclei. Here we use values for
$\tau_c$ in the range $ 0.1~{\rm fm/c} < \tau_c < 5$~fm/c,  and consider as upper bound the value $\tau_c = 10$~fm/c, 
which is approximately the lifetime of the medium produced at the RHIC and LHC. The cyclotron time $\tau_B = \om/eB$ 
is a momentum dependent quantity. For a given momentum, $\tau_B$ decreases with $B$ and $T$; the decrease with $B$ comes mainly 
from the trivial $1/eB$ factor, whereas the decrease in $T$ comes from $M$ in $\om = (\vp^2 + M^2)^{1/2}$. But $\tau_B$ 
is integrated over a momentum range bounded by the thermal distribution~$f_0(\om)$. To assess the 
effect of this integration, we have defined the following average: 
\be
\langle \tau_B \rangle = \frac{\int {d^3\vp} \, \tau_B \, f_0(\om)} {\int{d^3\vp} \, f_0(\om)} 
=  \frac{1}{\int{d^3\vp} \, f_0(\om)} \int {d^3\vp} \, \frac{\om}{eB} \, f_0(\om),
\label{av-tau}
\ee
with $\om = (\vp^2 + M^2)^{1/2}$. The temperature dependence of $\tau_B(B,T)$ is depicted in
Fig.~\ref{tB_T}(b). This average increases with~$T$ despite the decrease of $M$ with $T$. This is
because $f_0(\om)$ dominates the temperature dependence of the integrand; although the numerator and 
denominator in Eq.~(\ref{av-tau}) both grow with~$T$, the numerator grows faster.
This feature has important consequences for the temperature dependence of the anisotropy of the 
conductivity. The vertical arrow in Fig.~\ref{tB_T}(b) indicates the direction of growth of
$\langle \tau_B\rangle$ with $B$ and identifies regions of weak ($\tau_c < \tau_B)$ and strong
($\tau_c > \tau_B)$ magnetic field.

%%%%%%%%%%%%%%%%%%%%%%%%%%%%%%%%%%%%%%%%%%%
\begin{figure}[htb]
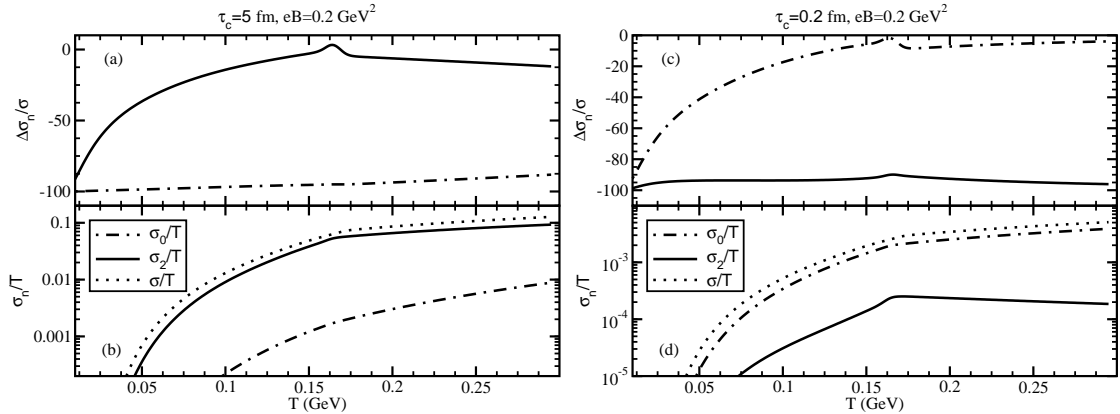

\begin{center}
\includegraphics[scale=0.3]{fig2a.eps}
\includegraphics[scale=0.3]{fig2b.eps}
\caption{Temperature dependence of the difference (\%) $\Delta \sigma_n/\sigma = (\sigma_n - \sigma)/\sigma$ and of $\sigma_n/T$, corresponding to the
three components of electrical conductivities for $eB = 0.2~{\rm GeV}^2$ and two values of the relaxation time,
$\tau_c = 5~{\rm fm/c}$ (left) and $\tau_c = 0.2~{\rm fm/c}$. Here the classical expressions for the $\sigma_n$, 
Eq.~(\ref{classical}), are used.} 
\label{el_T}
\end{center}
\end{figure}
%%%%%%%%%%%%%%%%%%%%%%%%%%%%%%%%%%%%%%%%%%%%%

Next, we present our results for the electrical conductivity{\textemdash}we recall the $B$ is taken in the 
$z$~direction. We start with the classical results. We recall that for zero magnetic field, the electrical conductivity tensor 
is diagonal, with all the diagonal elements being equal~to~$\sigma$, given by Eq.~(\ref{sigma_B0}). When $B\neq 0$, 
there is anisotropy in the conductivity, as indicated by Eqs.~(\ref{cond_ne1}) and (\ref{cond_ne2}). To quantify the 
anisotropy, we define the differences $\Delta \sigma_n/\sigma = (\sigma_n - \sigma)/\sigma$. In Fig.~\ref{el_T}, 
we present results for the temperature dependence of $\Delta \sigma_n/\sigma$ and of $\sigma_n/T$. The results 
are for $eB=0.2~{\rm GeV}^2$ and two values for the relaxation time, 
$\tau_c = 5~{\rm fm/c}$ and $\tau_c = 0.2~{\rm fm/c}$. For this value of $B$ and within the temperature range shown 
in the figure, $\tau_c = 5~{\rm fm/c}$ characterizes a strong magnetic field whereas $\tau_c = 0.2~{\rm fm/c}$
characterizes a weak magnetic field. It can be seen from Fig.~\ref{el_T} that $\sigma_n$ increase with the 
temperature having a kink at $T \simeq T_{\rm pc}$, after which the rate of increase is substantially reduced. 
This kink turns into a strong peak when using a $T$ and $B$ dependent $\tau_c$, as we discuss
at the end of this section. We conclude that the electrical conductivity becomes almost
independent of $T$ for $T > T_{\rm pc}$, when the constituent quark mass stops changing with $T$, since then $M \approx m$. 
In addition, $\sigma^{xx} = \sigma^{yy} < \sigma^{zz}$ 
for $\tau_c = 5~{\rm fm/c}$ (strong field), whereas $\sigma^{xx} = \sigma^{yy} \approx \sigma^{zz}$ for 
$\tau_c = 0.2~{\rm fm/c}$ (weak field), as expected. 

%%%%%%%%%%%%%%%%%%%%%%%%%%%%%%%%%%%%%%%%%%%%%
\begin{figure}[htb]
\begin{center}
\includegraphics[scale=0.3]{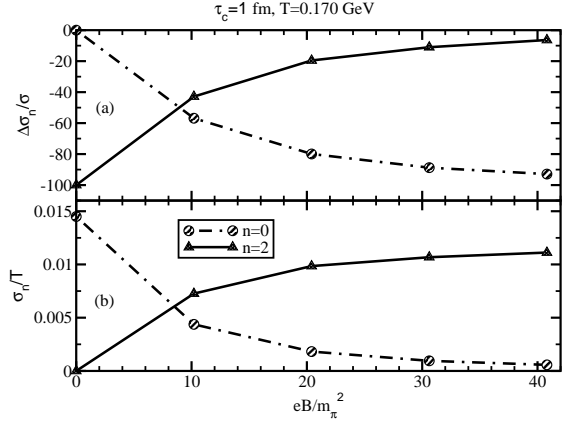} %{el_eB.eps}
\caption{Magnetic field dependence of the difference (\%) $\Delta \sigma_n/\sigma = (\sigma_n - \sigma)/\sigma$ and the
three components of electrical conductivities for $T = 0.170~{\rm GeV}$ and $\tau_c = 1~{\rm fm/c}$.
Here the classical expressions for the $\sigma_n$, Eq.~(\ref{classical}), are used.} 
\label{el_eB}
\end{center}
\end{figure}
%%%%%%%%%%%%%%%%%%%%%%%%%%%%%%%%%%%%%%%%%%%%%

Interesting insight is gained by examining the conductivity as a function of the magnetic field at
the pseudo-critical temperature. Fig.~\ref{el_eB} displays the $B$ dependence of $\Delta\sigma_n/\sigma$ 
and $\sigma_n/T$ at $T = T_{\rm pc} = 0.170~{\rm GeV}$, for $\tau_c = 1~{\rm fm/c}$. The anisotropy is enhanced
as $B$ increases. 
% %
% The Hall conductivity, $\sigma^{xy} = - \sigma^{xy} = \sigma_1$, presents a
% peak at $eB \simeq 10 m^2_\pi$. A peak in the Hall conductivity at some value of $B$ is easily 
% understood by using the simplified Drude relation~\cite{Drude_B1,Drude_B2,Feng_cond} for the conductivity:
% %
% \be
% \sigma_n = \left(\frac{q^2_f n_f \tau_c}{M}\right)\left[\frac{(\tau_c/\tau_B)^n}{1+(\tau_c/\tau_B)^2}\right],
% \ee
% %
% where $q_f$ and $n_f$ are respectively the quark electric charge and density which, together with the quark mass $M$ 
% are assumed to be independent of $B$. One obtains $d\sigma_1/dB = 0$ for $eB = M/\tau_c$. That is, there is 
% maximum in $\sigma_1$ when $\tau_B = {M}/{eB} = \tau_c$. On the other hand, $d\sigma_{0,2}/dB=0$ only at $B=0$. 
% %

The results discussed up to here used the classical expression for the conductivity, expected to
provide a reasonable approximation for weak magnetic fields. The quantum
description of charge transport leads to one-dimensional dynamics and quantization of the density of
states in terms of Landau levels in a plane perpendicular to the magnetic field. Taking the magnetic field 
in the $z$ direction, the longitudinal conductivity $\sigma^{zz}$ is then given by Eq.~(\ref{Lsig_QM}),
instead of $\sigma^{zz} = \sigma_0 + \sigma_2 = \sigma$, with $\sigma$ given by Eq.~(\ref{sigma_B0}). 
In the following we have compared the classical and quantum predictions of the model, and also examined the 
validity of truncating the density of states to the lowest Landau level (LLL). 

%%%%%%%%%%%%%%%%%%%%%%%%%%%%%%%%%%%%%%%%%%%%%%
\begin{figure}
\begin{center}
\includegraphics[scale=0.6]{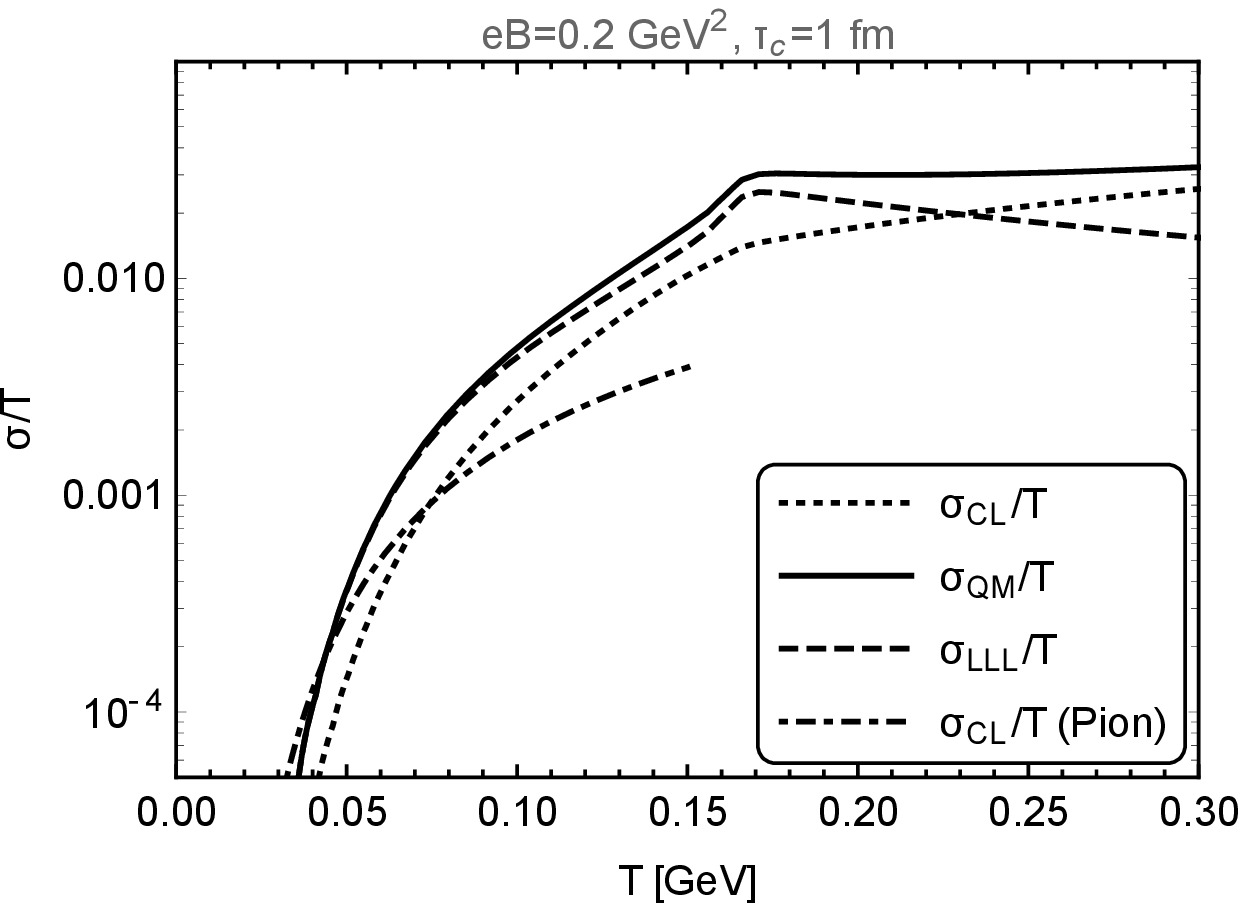} %{elzz_T.eps}
\hspace{0.2cm}
\includegraphics[scale=0.58]{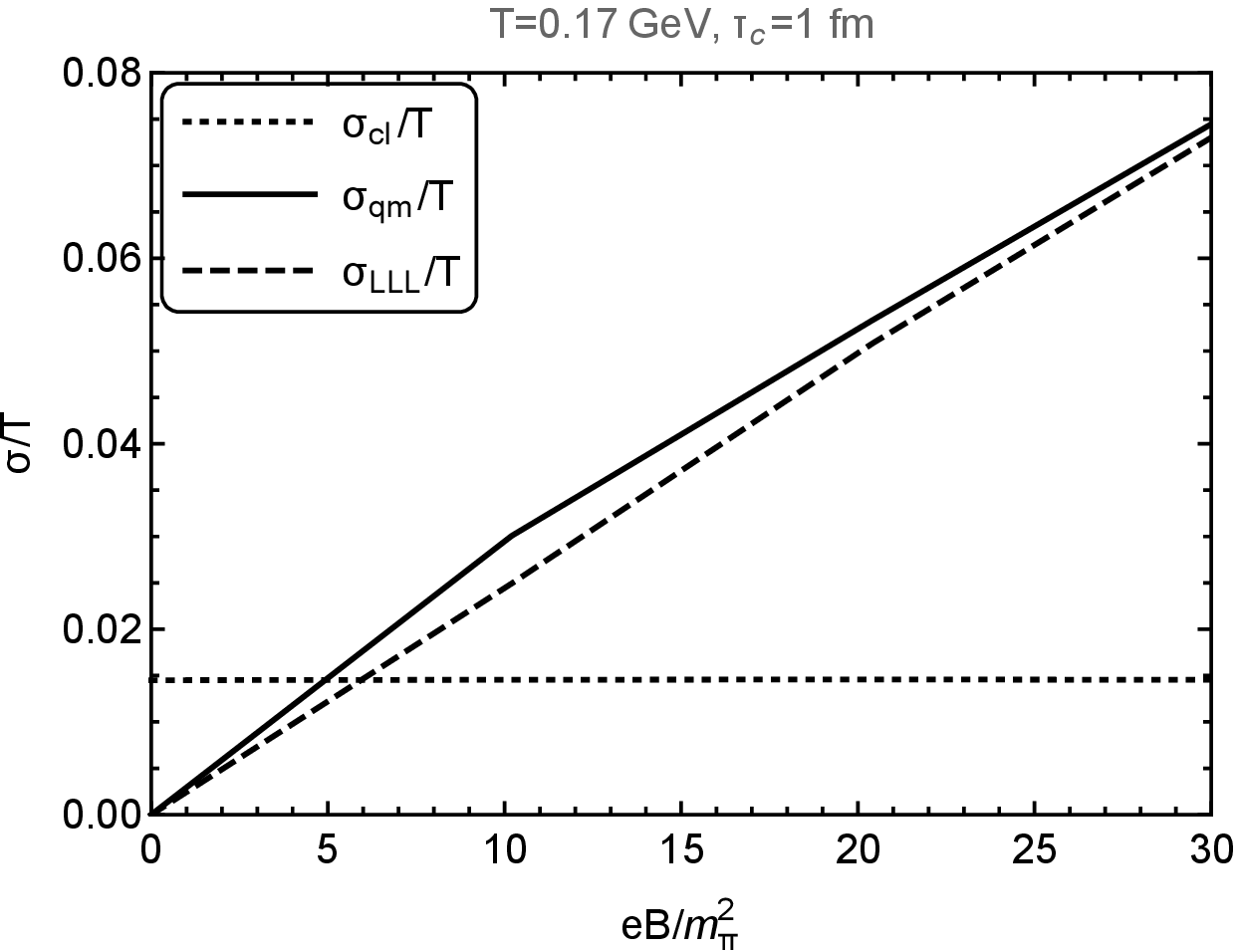} %{elzz_eB.eps}
\caption{Temperature (left panel) and magnetic field (right panel) dependent 
longitudinal electrical conductivity $\sigma^{zz}$. (For quark matter) dotted curve: classical, Eq.~(\ref{cond_ne2});
solid curve: quantum, Eq.~(\ref{Lsig_QM}), using $100$ Landau levels; dashed curve: LLL approximation. (For pion matter)
dash-dotted curve: classical, Eq.~(\ref{cond_ne2}).} 
\label{elzz}
\end{center}
\end{figure}
%%%%%%%%%%%%%%%%%%%%%%%%%%%%%%%%%%%%%%%%%%%%%%

The classical and quantum results for $\sigma^{zz}$ are shown in Fig.~\ref{elzz}. In the left panel we plotted 
$\sigma^{zz}$ as function of $T$ for $eB = 0.2~{\rm GeV}$, and on the right panel as a function of $B$ for 
$T = T_{\rm pc} = 0.170~{\rm GeV}$. All results are for $\tau_c = 1~{\rm fm/c}$. The classical and quantum 
results disagree in the entire temperature range shown in the figure, but the difference between them 
gets smaller as the temperature increases, as it should. In addition, the LLL approximation is good only for temperatures
below 0.1~GeV. Regarding the magnetic field dependence of $\sigma^{zz}$, the classical and quantum predictions
are completely different; this is expected, as the Lorentz force is ineffective along 
the direction of the magnetic field. The only effect of the magnetic field is via the mass $M$, 
but at $T=0.170$~GeV, one already has $M ~\simeq m$.  The truncation to the LLL is a poor approximation, 
particularly for values of $B$ in the region $6m_\pi^2<eB<10 m_\pi^2$. We find that quantum curve in right panel of Fig.~\ref{elzz} approximately proportional to $eB$ and cross the horizontal classical curve around $eB=5m_\pi^2$. So beyond the $eB=5m_\pi^2$ (for $T=0.170$ GeV, $\tau_c=1$ fm), quantum picture is basically revealing but within $eB=0-5m_\pi^2$ classical picture will be our matter of interest. Actually, due to numerical drawback
% {\footnote{Actually we have generated the results for 4 points of magnetic field ($eB/m^2_\pi=0, 10, 20, 30$ ) in right panel of Fig4. Now for latter three points, the convergence of results are established by choosing enough values of Landau level. One can understand that as magnetic field will decrease, values of maximum Landau level $l_{max}$ (which is numerically play role of infinity) will increase. And when eB reach to zero, then for getting trustable results, we have to choose infinite $l_{max}$, which is impossible.
% }}
we have not properly generated quantum results within $eB=0-5m_\pi^2$, which is ultimately supposed to merge with classical results. In general we can roughly assume low $T$ and/or high $eB$ as quantum zone and remaining part as classical zone.     

It is important to point out that our finding, i.e. the longitudinal conductivity ($\sigma^{zz}$) 
increases with~$B$ whereas the transverse conductivity ($\sigma_{xx} = \sigma_{yy}$) decreases with~$B$ 
is in qualitative agreement with the most recent lattice simulations of full QCD of Ref.~\cite{Astrakhantsev:2019zkr} 
(this reference presents results for $T = 0.2$~GeV and $T = 0.25$~GeV).

At low temperatures, the relevant degrees of freedom of the medium are hadrons. We make a comparison between 
the low temperature NJL results with those expected employing hadronic degrees of freedom; here we restrict the hadronic phase to pionic matter. Using the classical 
expression of longitudinal conductivity given in Eq.~(\ref{sig_cl}), we have obtained the conductivity for pionic 
matter. The required modifications in Eq.~(\ref{sig_cl}) are the transported electric charge and mass and the
the distribution function:  $\sum q_f^2$ is replaced by $2e^2$, and the $T$ and $B$ dependent quark mass 
is replaced by a constant 140 MeV pion mass, and $f_0$ will be the Bose-Einstein distribution instead of Fermi-Dirac distribution.
%(with zero chemical potential, since we are not interested in. 
The results are shown shown by dash-dotted line in the left 
panel of Fig.~(\ref{elzz}). We notice qualitatively similar temperature dependences of $\sigma_{zz}$ for
both quark and pionic matter at low temperatures.
We find that at $0<T<0.070$ GeV, pion component is lager than quark component but an opposite trend is found at $0.070<T<0.170$ GeV.
To realize actual mapping of NJL model with hadronic phase, we should go through their normalized value. From Fig.~\ref{elzz}, one can notice that
$\sigma/T=0.0018-0.0047$ for pion medium and $\sigma/T=0.0028-0.014$ for NJL based quark medium within $T=0.1-0.17$ GeV. However, if we normalize their charge factors, then we will get the ranges:
\bea
\frac{\sigma}{T}\Big(\frac{1}{2e^2}\Big)&=&\frac{0.0009}{e^2}-\frac{0.0023}{e^2},~{\rm (for ~pion~medium)}
\nn\\
\frac{\sigma}{T}\Big(\frac{9}{5e^2}\Big)&=&\frac{0.005}{e^2}-\frac{0.025}{e^2},~{\rm (for ~NJL-based~quark~medium)}~.
\eea
Now, we noticed that by normalizing the charge factor, longitudinal conductivity of pion medium
are lower than NJL-based quark medium in entire hadronic temperature $0<T<0.170$ GeV.
If one follow hadron resonance gas (HRG) model~\cite{HRGB} and add higher mass resonances with pion component, then hadronic matter conductivity might be close to the conductivity of NJL-based quark medium.
For thermodynamical quantities like pressure, energy density etc. the approximate equivalence of NJL model, LQCD and HRG model in low temperature ($T=0.1-0.17$ GeV) is quite well known fact.

%%%%%%%%%%%%%%%%%%%%%%%%%%%%%%%%%%%%%%%%%%%%%%%%%%%%%%%%%%
\begin{figure}
\begin{center}
\includegraphics[scale=0.3]{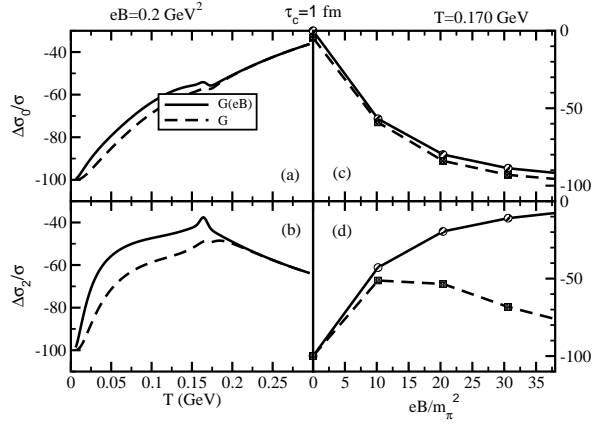} %{MCIMC.eps}
\caption{Effect of inverse magnetic catalysis (IMC) on the electrical conductivity.
The solid lines are obtained with a constituent quark mass $M$ calculated with the NJL coupling 
$G = G(eB,T)$ and the dashed lines with $G=G(0,0)=4.6311$ GeV$^{-2}$.} 
\label{MCIMC}
\end{center}
\end{figure}
%%%%%%%%%%%%%%%%%%%%%%%%%%%%%%%%%%%%%%%%%%%%%%%%%%

Finally, we examined the significance of IMC on the electrical conductivity. Fig.~\ref{MCIMC} shows two sets of 
results for the $T-$ and $B-$dependence of $\Delta \sigma_n/\sigma$: one is calculated with a constituent quark 
mass~$M$ with the $G(eB,T)$ coupling of Eq.~(\ref{GBT}), and  the other is calculated with a $G$ fixed to 
$G = G(eB,T)=4.6311\,$ GeV$^{-2} $. We recall that the NJL model describes IMC at $T \sim T_{\rm pc} = 0.170$~GeV 
only when $G = G(eB,T)$, otherwise it gives magnetic catalysis (MC) in the entire range of temperatures shown 
in the figure. IMC produces a kink in the $\Delta\sigma_n$ at $T \simeq T_{\rm pc}$, 
whereas MC produces featureless $\Delta\sigma_n$. The physical origin of this kink structure can be understood 
by comparing the constituent quark mass $M$ calculated with
a $G=G(eB,T)$ and with a $G$~fixed. The comparison is better made by defining the ratio  
$\Delta M/M_{0}$, where $\Delta M = M(B) - M_0$ and $M_0=M(B=0)$. Fig.~\ref{M_MCIMC} displays the 
comparison: the kink is due to a rapid inversion of the temperature dependence of $M$ when $G=G(eB,T)$. 
In the following, we show that this kink in the conductivity becomes a prominent 
peak when using a $T$ and $B$ dependent $\tau_c$.

%%%%%%%%%%%%%%%%%%%%%%%%%%%%%%%%%%%%%%%%%%%%%%
\begin{figure}[htb]
\begin{center}
\includegraphics[scale=0.3]{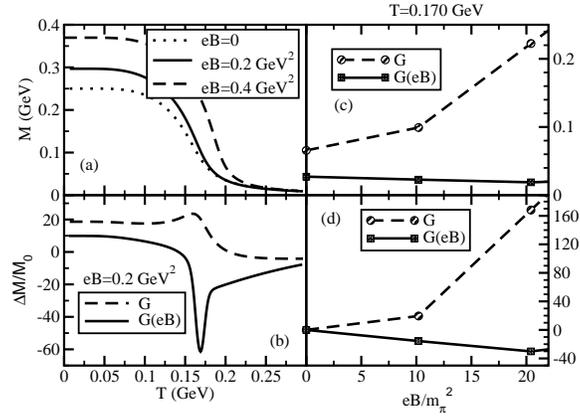} %{M_MCIMC.eps}
\caption{(a): Constituent quark mass for fixed constant coupling $G=G(0,0)=4.6311$ GeV$^{-2}$.
 (c) Constituent quark mass for $G = G(eB,T)$.
(b) and (d): $\Delta M/M_0$, where $\Delta M = M(B) - M_0$ and $M_0=M(B=0)$.   } 
\label{M_MCIMC}
\end{center}
\end{figure}
%%%%%%%%%%%%%%%%%%%%%%%%%%%%%%%%%%%%%%%%%%%%

The results above are for fixed values of the relaxation time $\tau_c$. Next, we examine the impact 
of a $T$ and $B$ dependent $\tau_c$ on the longitudinal 
conductivity. A a $T$ and $B$ dependent $\tau_c$ can be calculated within the NJL model: 
for $B=0$, the temperature dependence of $\tau_c$ was computed in 
Refs.~\cite{G_CAPSS,G_IFT,Weise2,LKW} 
using quark-$\pi$ and quark-$\sigma$ loops, in Refs.~\cite{klevansky,klevansky2,
Redlich_NPA,Marty,Deb} it was computed using quark-quark scattering diagrams
with $\pi$ and $\sigma$ exchanges, and in Ref.~\cite{G_IFT2} it was computed 
using both quark-meson loops and quark-quark scatterings. The extension of those
calculations to $B \neq 0$ is technically difficult,
but it can be simplified employing quark-quark scattering processes
from the contact interactions of the NJL model~\cite{G_shear_NJLB}. 
Here we follow this approach using a fixed $G$ and a $G(T,eB)$ coupling as 
above. 

The relaxation time is given by~\cite{G_shear_NJLB}:
\bea
&&\frac{1}{\tau_c(T,eB)}=\Gamma_{c}(T,eB)=\frac{\sum_{k=0}^\infty 
\alpha_k \, \frac{|q_f|B}{2\pi} \int \frac{dp_a^z}{2\pi} \, \Gamma(T,eB,\vp_a) f_{0}(\omega_a)} 
{\sum_{k=0}^\infty \alpha_k\frac{|q_f|B}{2\pi}\int \frac{dp_a^z}{2\pi} f_0(\omega_a)},
\label{GQ_T}
\eea
where $\om_a$ is given in Eq.~(\ref{omega_f}), $\Gamma(T,eB)$ is the momentum average of
the the collisional width (collisional frequency) of probe particle with momentum $\vp_a$,
$\Gamma(T,eB,\vp_a)$. The collisional width $\Gamma(T,eB,\vp_a)$ is given in terms of quark-quark 
scattering cross-sections as
\bea
&&\Gamma_c(T,eB,\vp_a)=\frac{1}{\tau_c(T,eB,\vp_a)}
\nn\\
&&=\sum_{b}\sum_{k=0}^\infty \alpha_k\frac{|q_f|B}{2\pi} 
\int \frac{dp_{b z}}{2\pi} \, \sigma_{ab}(T,eB,\vp_a,\vp_b) v_{ab}(T,eB,\vp_a,\vp_b) 
f_0(\omega_b) ,
\label{G_KTmu}
\eea
where 
\be
v_{ab}(T,eB,\vp_a,\vp_b) = \frac{\{(\om_a +\om_b)^2-4M_Q^2(T)\}^{1/2}(\om_a +\om_b)}{2\om_a\om_b},
\ee
is relative velocity.  The $\sigma_{ab}$ cross-sections are obtained~\cite{G_shear_NJLB} using
the standard quantum field theoretical relation of $2\rightarrow 2$ scattering as
\be
\sigma_{ab} = \frac{1}{16\pi (\om_a +\om_b)^2}{\overline{|M|_{ab}^2}} 
= \frac{(\om_a +\om_b)^2}{16\pi} \, G^2(eB,T) .
\ee
Hence, we have $T$ and $B$ dependence in the cross-sections through the $\om_a$ and $\om_b$, 
and also through the coupling $G(eB,T)$. Fig.~\ref{fig8}(a) shows the temperature dependence 
of $\tau_c$ for different values of~$B$, for $G$ fixed and $G(eB,T)$. Fig.~\ref{fig8}(b) shows 
the corresponding temperature dependence of longitudinal conductivity. For numerical simplicity, 
both of these plots are for the lowest Landau level. The interpretation of the results is 
straightforward, as we discuss next.

\begin{figure}
\begin{center}
\includegraphics[scale=0.8]{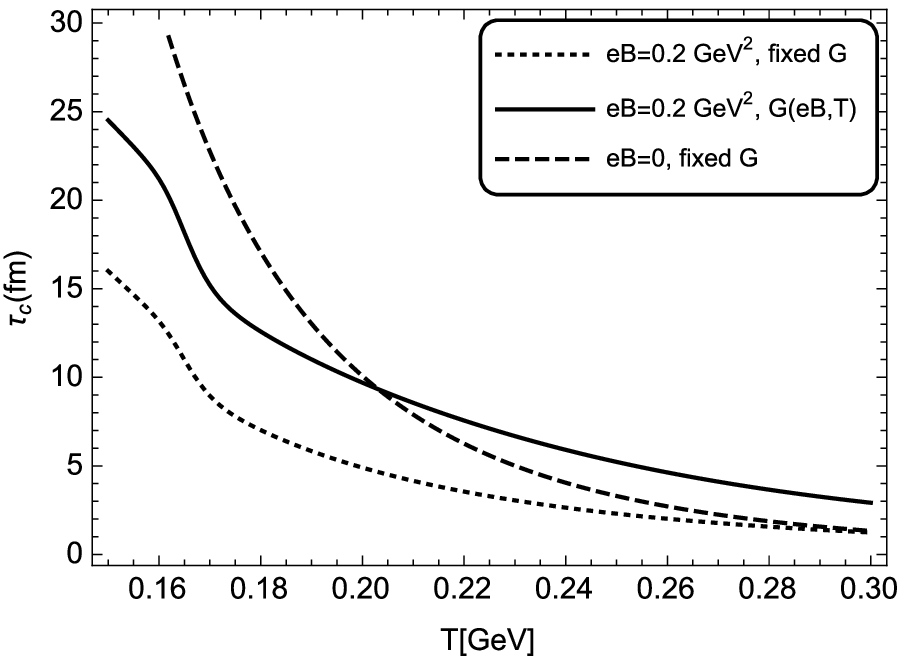} \hspace{0.5cm}\includegraphics[scale=0.58]{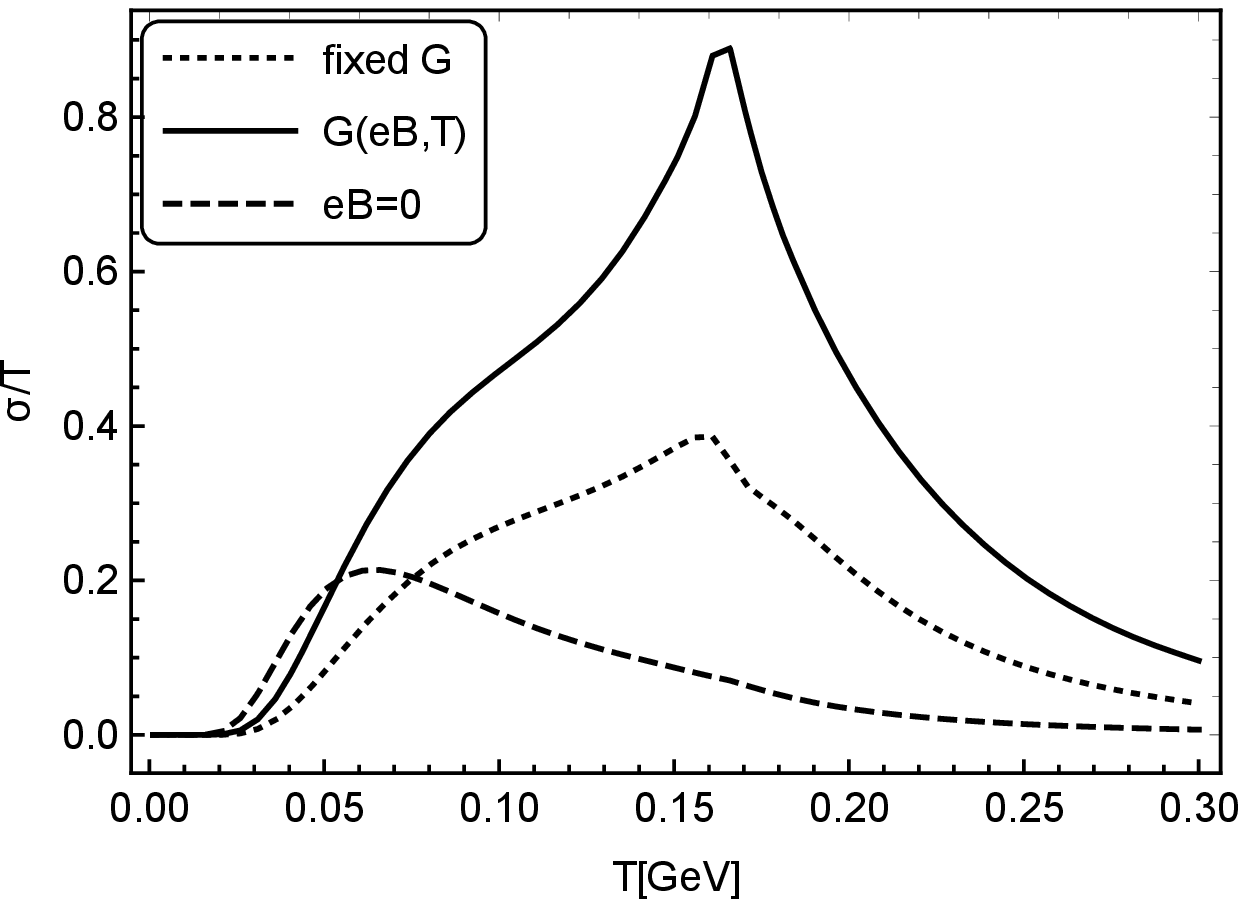} 
\caption{
%{\color{red}I suggest to draw the figures in the same style as the %others: square-box
%legends, frame lines a bit thicker}. 
Temperature dependence of $\tau_c$ (left panel) and $\sigma/T$ (right panel) for 
different values of $eB$ and fixed (dashed lines) and $T$ and $B$ dependent (solid and dotted 
lines) NJL coupling constants.} 
\label{fig8}
\end{center}
\end{figure}

Qualitatively, Eq.~(\ref{G_KTmu}) means
$\tau_c^{-1}=\sigma_{ab}\times v_{ab}\times\rho$, where $\rho$ is number density.
Being inversely proportional to~$\rho$, $\tau_c$ decreases with the increase 
of $T$. For massless quarks, a Stefan-Boltzmann type density follows a $T^3-$dependence, 
which is definitely modified for a $T$-dependent constituent quark mass. Now, for nonzero
magnetic field, the phase-space distribution $f_0(\om)$ in general is 
enhanced, as seen in the quark condensate and constituent quark mass. This means that 
$\tau_c$ is reduced due to the magnetic field. This reduction of~$\tau_c$ can be seen 
in Fig.~\ref{fig8}(a) by comparing the dashed and dotted lines. We note that that 
these two curves merge at high $T$, when the quark condensate becomes very small. 
These results are for a fixed value of the coupling~$G$, the model describes magnetic 
catalysis only. The effect of inverse magnetic catalysis on $\tau_c$, obtained with a 
$G(eB,T)$ coupling constant, is shown  by the solid line in Fig.~\ref{fig8}(a): a rapid
fall of $\tau_c$ at $T_{\rm pc}$. In addition, there is an inversion of the temperature 
dependence at low (high) $T$, $\tau_c$ is smaller (larger) than the corresponding values 
at $B=0$. This rapid decrease in $\tau_c$ around the pseudocritical temperature leads 
to a much stronger peak in the conductivity than the one observed at with a fixed value 
of~$G$, as shown in Fig.~\ref{fig8}(b). This prediction of the model can be tested by 
lattice QCD simulations.

\section{Summary and Perspectives} 
\label{sec:summary}

Relativistic heavy collisions can produce strong magnetic fields. Strong magnetic fields have striking effects 
on properties of the hot quark matter created in such collisions. Measurability of many of the effects 
depend on the duration of the field produced in a collision. A key physical property determining the duration of the 
field is the electrical conductivity of the medium: the conductivity is responsible for the induction of electric 
currents which in turn can produce magnetic fields that can last while the systems exists. Simulations of the field
dynamics involves solving relativistic magnetohydrodynamics equations which, for a realistic quantification require 
the temperature and magnetic field dependence of the electrical conductivity (and, of course, of other transport coefficients). 
We focused on the implications for the conductivity of changes induced by a magnetic field on the effective 
quark masses, with special interest on the importance of magnetic catalysis (MC) and inverse magnetic catalysis 
(IMC){\textemdash}while MC occurs at low temperatures, IMC occurs at temperatures close to the pseudo-critical 
temperature ($T_{\rm pc}$) of the QCD phase transition. We employed a quasi-particle model for the medium, implemented 
by a Nambu--Jona-Lasinio model with a temperature- and magnetic-field-dependent coupling constant adjusted to lattice 
QCD data on MC and IMC. One of our main finding was that while the longitudinal conductivity ($\sigma^{zz}$) increases 
with~$B$, the transverse component ($\sigma^{xx} = \sigma^{yy}$) decreases with~$B$, in qualitative agreement with the 
very recent lattice QCD results of Ref.~\cite{Astrakhantsev:2019zkr}. Moreover, 
our study revealed that IMC leaves a distinctive signal in all components of conductivity, a kink at $T_{\rm pc}$.
In addition, IMC makes the peak more prominent in the quark and antiquark contributions to $\sigma^{xy} = - \sigma^{yx}$ 
at $eB \simeq 10 m^2_\pi$. Such a peak appears when the two time scales controlling the conductivity, i.e. the 
relaxation time $\tau_c$ and
the inverse of the cyclotron frequency $\tau_B$ are comparable. This feature of IMC on these components is a prediction of
the model and is testable with lattice QCD simulations. Additional findings in our study were: (1)~for a fixed value of $B$, 
all components of the conductivity increase with the temperature, (2)~the anisotropy in the conductivity increases with~$B$, (3)~quantum effects leading to phase-space Landau-level quantization increase the longitudinal conductivity, 
and (4)~truncation to the lowest Landau level gives a poor approximation for temperatures close to $T_{\rm pc}$.

Our study adds to the existing body of work on the hot quark matter electrical conductivity by incorporating 
nontrivial temperature and magnetic field effects on dynamical mass generation, such as MC and IMC.
The results are useful for studies employing magnetohydrodynamics simulations of heavy-ion collisions and,
on a wider perspective, they give insight on recent lattice QCD results on the electrical conductivity of
magnetized hot quark matter.

\vspace{1.0cm}
%%%%%%%%%%%%%%%%%%%%%%%%%%%%%%%%%%%%%%%%%%%%%%%%%%%%%
%
{\bf ACKNOWLEDGMENTS}
S.G. and J.D. acknowledge to IIT-Bhilai research facilities, funded by Ministry of Human Resource Development (MHRD), 
Government of India. Work partially supported by Conselho Nacional de Desenvolvimento Cient\'{\i}fico 
e Tecnol\'ogico - CNPq, Grants. no. 304758/2017-5 (R.L.S.F), 305894/2009-9 (G.K.), and 464898/2014-5(G.K) 
(INCT F\'{\i}sica Nuclear e Aplica\c{c}\~oes), and Funda\c{c}\~ao de Amparo \`a Pesquisa do Estado do  
Rio Grande do Sul - FAPERGS, Grant No. 19/2551-0000690-0 (R.L.S.F.), and 
Funda\c{c}\~ao de Amparo \`a Pesquisa do Estado de S\~ao Paulo - FAPESP, 
Grant No. 2013/01907-0 (G.K.) and Coordena\c c\~ao de Aperfei\c coamento de Pessoal  de N\'ivel Superior (CAPES) (A.B.) - Brasil (CAPES)- Finance Code 001.

\end{document}